\documentclass[sigconf]{acmart}

\usepackage{booktabs} 
\usepackage{pgfplots} 
\pgfplotsset{compat=newest}
\usepgfplotslibrary{groupplots}

\usepackage[ruled]{algorithm2e} 
 
\SetAlFnt{\small}
\SetAlCapFnt{\small}

\SetAlCapNameFnt{\small} 
\SetAlCapHSkip{0pt} 
\IncMargin{-\parindent}

\setcopyright{rightsretained} 

\begin{document} 
\title{Learning to Rank Using Localized Geometric Mean Metrics}

\author{Yuxin Su}
\affiliation{
	\department{Department of Computer Science and Engineering}
	\institution{The Chinese University of Hong Kong}
	\streetaddress{Shatin, N.T., Hong Kong}
}
\email{yxsu@cse.cuhk.edu.hk}
	
\author{Irwin King}
\affiliation{
	\department{Department of Computer Science and Engineering}
	\institution{The Chinese University of Hong Kong}
	\streetaddress{Shatin, N.T., Hong Kong}
}
\email{king@cse.cuhk.edu.hk}
	
\author{Michael Lyu}
\affiliation{
	\department{Department of Computer Science and Engineering}
	\institution{The Chinese University of Hong Kong}
	\streetaddress{Shatin, N.T., Hong Kong}
}
\email{lyu@cse.cuhk.edu.hk}

\renewcommand{\shortauthors}{Su, King and Lyu}

\begin{abstract} 
Many learning-to-rank~(LtR) algorithms focus on query-independent model, in which query and document do not lie in the same feature space, and the rankers rely on the feature ensemble about query-document pair instead of the similarity between query instance and documents.
However, existing algorithms do not consider local structures in query-document feature space, and are fragile to irrelevant noise features. 
In this paper, we propose a novel Riemannian metric learning algorithm to capture the local structures and develop a robust LtR algorithm. 
First, we design a concept called \textit{ideal candidate document} to introduce metric learning algorithm to query-independent model. 
Previous metric learning algorithms aiming to find an optimal metric space are only suitable for query-dependent model, in which query instance and documents belong to the same feature space and the similarity is directly computed from the metric space.
Then we extend the new and extremely fast global Geometric Mean Metric Learning (GMML) algorithm to develop a localized GMML, namely L-GMML.
Based on the combination of local learned metrics, we employ the popular Normalized Discounted Cumulative Gain~(NDCG) scorer and Weighted Approximate Rank Pairwise~(WARP) loss to optimize the \textit{ideal candidate document} for each query candidate set. 
Finally, we can quickly evaluate all candidates via the similarity between the \textit{ideal candidate document} and other candidates. 
By leveraging the ability of metric learning algorithms to describe the complex structural information, our approach gives us a principled and efficient way to perform LtR tasks. 
The experiments on real-world datasets demonstrate that our proposed L-GMML algorithm outperforms the state-of-the-art metric learning to rank methods and the stylish query-independent LtR algorithms regarding accuracy and computational efficiency.
\end{abstract} %

\copyrightyear{2017}
\acmYear{2017}
\setcopyright{acmcopyright}
\acmConference{SIGIR '17}{August 07-11, 2017}{Shinjuku, Tokyo, Japan}
\acmPrice{15.00}
\acmDOI{http://dx.doi.org/10.1145/3077136.3080828}
\acmISBN{978-1-4503-5022-8/17/08}




\maketitle

\section{Introduction}

In many information retrieval systems, especially Web search, users expect to obtain the most relevant documents according to users' query phrase or document. 
This task is technically formulated as a ranking problem. 
Most of the Web search engines exploit this ranking task based on learning-to-rank~(LtR) techniques~\cite{2009_ftir_t.liu_learning}. 
In the LtR framework, a machine learning algorithm is typically employed to derive a ranking model about document collection from a training subset of documents with labels or partial order. 
After the supervised or semi-supervised learning procedures, the ranking model is expected to retrieval top-$k$ (ordered) relevant documents from the candidate collection when a query is given.

In practice, search engines develop the LtR model in two stages:
(i) candidate retrieval and (ii) candidate re-ranking~\cite{ 2015_sigir_c.lucchese_speeding}. 
In the first stage, search engine retrieves from the inverted document repository a sufficiently large set of relevant candidate documents $\mathcal{D}_{q}$ matching a user's query. 
It is used to avoid applying the ranking model to all documents possibly matching a user's query. 
This stage usually requires that the size of candidate set is much larger than the number of the relevant URLs to be included in the returned page. 
Based on the candidate document set $\mathcal{D}_{q}$ obtained in the first stage, Web search engines reformulate the documents with features extracted from the query-document pair and hide query features, then employ the LtR model without the dependency of query instance to score and re-rank the document collection $\mathcal{D}_{q}$.
Finally, search engines return the top-$k$ documents to the user.

In Web search engine, the time-budget of this two-stage framework is usually
limited. 
Therefore, strongly motivated by the time budget consideration, the current two most efficient and the state-of-the-art methods are based on the additive ensemble of regression trees, namely Gradient-Boosted Regression Tree (GBRT)~\cite{ 2001_as_j.friedman_greedy} and $\lambda$-MART~\cite{ 2011_jmlr_ws_c.burges_learning}. 
These two kinds of methods are capable of meeting the time requirement with acceptable accuracy even when thousands of regression tree are evaluated for each document.
However, one of the drawbacks of this line of methods is that when the input samples contain an enormous amount of non-informative features, many methods fail to identify the most relevant features. 
Therefore, researchers are still trying to devise techniques and strategies to find a better way of combining features extracted from query-document pairs through discriminative training to accelerate the training process for document ranking without losing in quality~\cite{2011_sigir_y.ganjisaffar_bagging, 2012_icml_z.xu_greedy}.

Another perspective to the ranking problem is to seek the best similarity measurement and develop the corresponding efficient algorithm. 
These approach aims to optimize the accuracy in the first stage to find candidate documents or even directly return the top-$k$ documents with an order. 
Concerning accuracy, the similarity-based models for a ranking problem can be classified into three categories from the formulation of the loss function: pointwise, pairwise and listwise loss functions~\cite{ 2010_tr_c.burges_ranknet}. 
Practically, the pairwise loss function tends to be more efficient for training and have been widely adopted by large Web search engines~\cite{ 2010_tr_c.burges_ranknet}.

The pairwise similarity motivates that how to apply the classical metric learning or similar learning methods to the ranking problem~\cite{ 2010_jmlr_g.chechik_large}. 
The metric learning algorithms aim to find a better distance metric than Euclidean metric to measure the pairwise similarity. 
The advantage of such metric-learning-to-rank~\cite{ 2012_itaslp_b.mcfee_learning} framework has two folds: (1) the metrics often preserves the nearest neighborhood information, which is the perfect structure to conduct ranking; 
(2) a proper metric containing the structural information of the document collections in the document space is useful for reducing the over-fitting and improving the robustness to noise features~\cite{2013_ftml_b.kulis_metric}.
Therefore, the metric-learning-to-rank methods~\cite{ 2010_icml_b.mcfee_metric, 2013_icml_d.lim_robust, 2014_icml_d.lim_efficient} typically enjoy higher accuracy. 
Nevertheless, unfortunately, the disadvantage of metric-learning-to-rank also has two folds: (1) many metric learning algorithms~\cite{ 2007_icml_j.davis_information, 2009_jmlr_k.weinberger_distance} are degraded by its extremely high computational expense; 
(2) the similarity measurement is not suitable for LtR because we can not estimate the similarity between query and documents with features extracted from other domain knowledge.

In this paper, we focus on improving the ranking accuracy at the second stage in the search engine and attempt to provide a new query-independent model for LtR task. 
Different from the existing research on how to combine features extracted from other domains, we try to learn an optimal representation of these features via metric learning algorithm. 
To adopt query-dependent metric learning framework to a query-independent model, we propose a concept called \textit{ideal candidate document}, which represents a perfect match for a given query. 
With the help of this concept, we can quickly evaluate all candidate documents and sort them by calculating the distance based on the optimal metric space between the \textit{ideal candidate document} and other documents. 
Same with the query-dependent model, the shorter distance leads to a higher relevance to the query.

Since features from different domains generate local structure on the whole feature space, in order to preserve more local information and avoid overfitting, we develop a novel local metric learning framework for ranking with high efficiency and accuracy. 
Our localized metric learning algorithm extends from the state-of-the-art global metric learning algorithm called Geometric Mean Metric Learning (GMML)~\cite{ 2016_icml_p.zadeh_geometric}, and we apply Weighted Approximate Rank Pairwise~(WARP) loss to optimize the metric space around the ideal document from the combination of several anchor documents.

We summarize our main contributions as follows: 
\begin{itemize} 
	\item To the best of our knowledge, we are the first to extend geometric mean metric learning algorithm to a local metric learning approach in order to capture the local structures for LtR problem. 
	\item We propose a novel \textit{ideal candidate document} concept to transform metric-learning-to-rank framework from query-dependent model to query-independent model, which brings wider applications for metric learning and also improves the accuracy of classical LtR task. 
	\item We conduct extensive experiments to reveal that our method outperforms the state-of-the-art query-dependent metric-learning-to-rank algorithms and query-independent LtR methods both in the accuracy and the computational complexity.
\end{itemize}

\section{Preliminaries and Related Work}

Since our approach employs local metric learning algorithm to conduct the LtR task, two sets of previous work relate to our work:

\subsection{Learning to Rank}

In the information retrieval setting, a search engine maintains a collection of candidate examples $\mathcal{D}$. 
Given a query $q$, the search engine returns the top ranked subset of documents $\left\{ p\in\mathbb{R}^{d}\right\} \subset\mathcal{D}_{q}\subset\mathcal{D}$ from the collection with order, ranked by a specific ranking model $f_{q}(p)$.

According to the formulation of the loss function, the LtR methods are categories into three folds: (1) pointwise loss approach, (2) pairwise loss approach and (3) listwise loss approach.

For pointwise loss function, Li et al.~\cite{2008_nips_p.li_mcrank} cast the ranking problem to a multi-class classification problem. 
The training process relies on enough labeled information, which is not always easy to satisfy.
Pairwise loss approach such as RankNet~\cite{ 2005_icml_c.burges_learning}, RankBoost~\cite{ 2003_jmlr_y.freund_efficient} focus on the relative order, which is capable of being adapted to classification problem. 
In the listwise loss approach, a relevance label $l$ related with the query for ground truth is usually bound to the document $p$. 
Cao et al.~\cite{2007_icml_z.cao_learning} first propose to find the optimal permutation to minimize the listwise loss function. 
McFee~\cite{2010_icml_b.mcfee_metric} proposes a similar objective, but the different solution from the metric learning methods.

The majority of LtR methods follows listwise loss function.
Currently, the most popular methods~\cite{ 2016_sigir_c.clebson_generalized, 2001_as_j.friedman_greedy, 2011_jmlr_ws_c.burges_learning} come from the combination of an ensemble of trees like random forest and the boosting-like methods~\cite{2003_jmlr_y.freund_efficient}. 
Based on multiple decision trees, this kind of methods gains an accepted level of accuracy.

\subsection{Metric Learning}

The (squared) Mahalanobis distance, an extension of Euclidean distance, measures the distance between two points lie on the special linear space. 
It is defined as

\begin{equation}
d_{\mathbf{M}}\left(p_{1},p_{2}\right)=\left(p_{1}-p_{2}\right)^{T}\mathbf{M}\left(p_{1}-p_{2}\right),
\label{eq:global_m_distance} 
\end{equation} 

where $p_{1},p_{2}\in\mathbb{R}^{d}$ are input examples, $\mathbf{M}$ is a symmetric and positive semi-definite $d\times d$ matrix. 
When $\mathbf{M}=\mathbf{I}$, the Mahalanobis distance is equivalent to the Euclidean distance.

There are plenty of algorithms aiming at learning such metric by solving a semidefinite or a quadratic program~\cite{ 2009_jmlr_k.weinberger_distance, 2002_nips_e.xing_distance, 2012_jmlr_c.shen_positive}. 
Almost all the metric learning algorithms try to constrain the similar data points and to scatter those dissimilar data points. 
Early work like~\cite{2002_nips_e.xing_distance} formulates this problem as an optimization problem on the second-order cone, which is costly solvable. 
Davis et al.~\cite{ 2007_icml_j.davis_information}, Weinberger et al.~\cite{ 2009_jmlr_k.weinberger_distance} and Shen et al.~\cite{ 2012_jmlr_c.shen_positive} formulate different kinds of optimization problems, namely ITML, LMNN, BoostMetric respectively.
However, the common issue that their solutions are computationally expensive. 
Very recently, Zadeh et al.~\cite{2016_icml_p.zadeh_geometric} propose a new objective function and give the closed-form solution from the geometric domain. 
It is the most promising global metric learning method because of the computational speed several orders of magnitude faster than the widely used ITML and LMNN methods.

There are two different roadmaps to conduct the LtR tasks from the metric learning perspective: (1), McFee~\cite{ 2010_icml_b.mcfee_metric} and Lim et al.~\cite{ 2013_icml_d.lim_robust, 2014_icml_d.lim_efficient} learn global metric with Positive Semi-Definite (PSD) constraint on the metric parameter.
They belong to the application of the standard metric learning algorithm.
(2), Chechik et al.~\cite{ 2010_jmlr_g.chechik_large} and Liu et al.~\cite{ 2015_aistats_k.liu_similarity} remove PSD constraint and employ the bilinear model to measure the similarity between two data points. 
Usually, without PSD constraints, the bilinear model easily leads to over-fitting.
However, PSD constraint brings a tremendous amount of computation.

In most cases, global metric learning relies on a learned PSD matrix, which is not only computational expensive in high-dimensional case but not reasonable for retrieval ranking problem. 
In the LtR framework, the local similarity is far more important than the dissimilar information because we aim at ranking the relevant documents around a user's query. 
Therefore, several important local metric learning approaches are related to our work. 
Wang et al.~\cite{ 2012_nips_j.wang_parametric} parameterize the weight function of each data point. 
The approach enhances the model complexity but brings the extra computation. 
Hauberg et al.~\cite{ 2012_nips_s.hauberg_geometric} provide the theoretical analysis about the optimal weight function. 
However, the calculation of the geodesics is extremely expensive.

\section{Our Proposed Methodology}

In the LtR problem, a ranked list of the relevant documents is returned for a specific user\textquoteright s query.  
In this situation, we can assume without losing generality that all relevant documents should be closer to an unreal document than other irrelevant documents. 
This unreal document should be related to the query. 
Therefore, although the query instance is not accessible in the document feature space, we can still construct this unreal candidate document to represent the query in the feature space of the document.
In our paper, this unreal but perfect-matching document is named as the \textit{ideal candidate document}.

\begin{figure*} 
	\centering
	\includegraphics[width=0.8\textwidth]{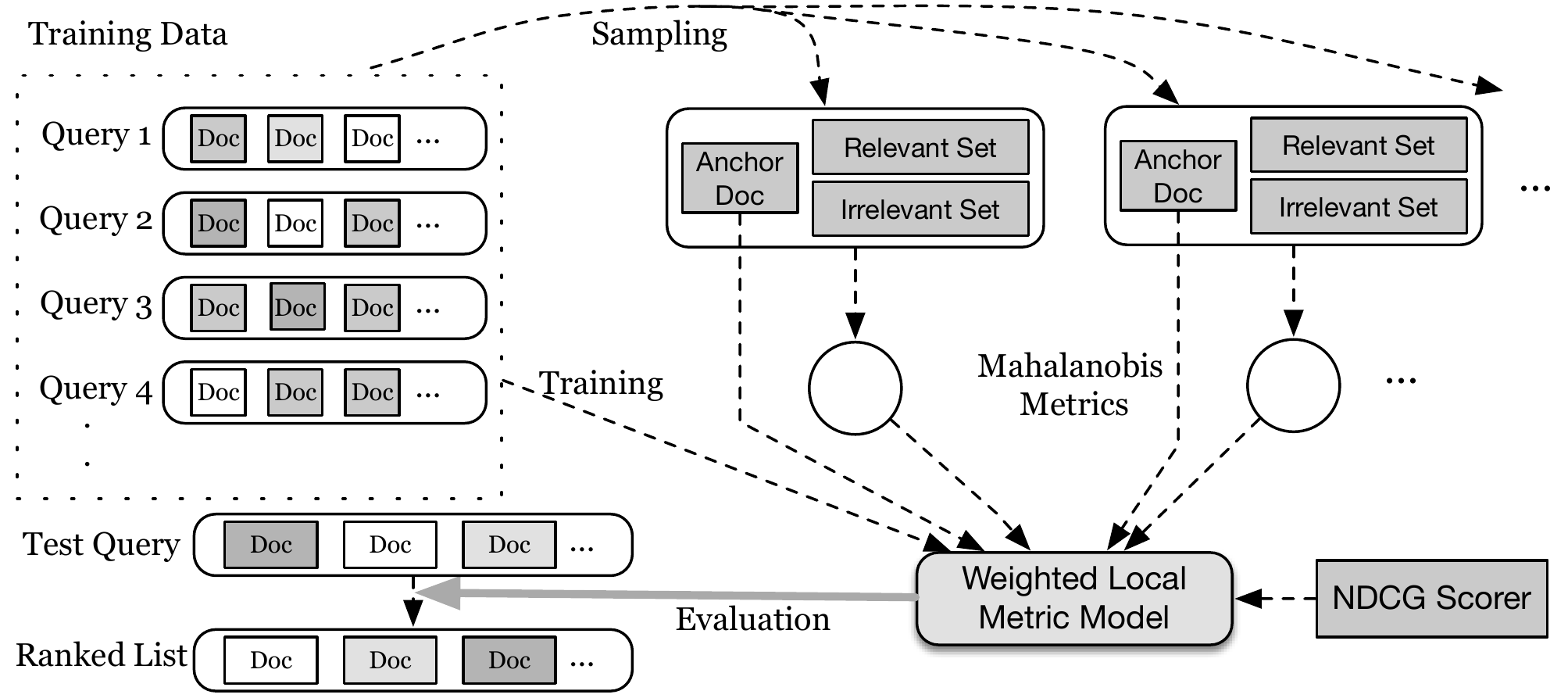}
	\caption{General framework of proposed L-GMML for ranking. Different gray
		levels in query test represent the relevant level of the document}
	\label{fig:framework} 
\end{figure*}

Usually, the indexed documents are assumed to be static, and the set of queries considered as input testing data is dynamic. 
This assumption allows us to transform the training documents to an another static representation, and model \textit{ideal candidate documents} for each query to a dynamic combination of static documents.

In our paper, we assume the documents including candidate documents and ideal documents lie on the surface of a Riemannian manifold. 
Then, we attempt to build the similarity measurement between documents on the geodesic lines in the Riemannian manifold.
Very often, a single linear metric $\mathbf{M}$ can not describe the whole surface of Riemannian manifold adequately. 
It reveals the inability of a single metric to model the complexity of the LtR problem. 
Furthermore, the discriminative features vary between different neighborhoods on the surface of the manifold. 
To address this limitation, researchers try to learn a set of local metrics representing the various regions of the surface.
In most cases, local metric learning algorithms will generate a local metric for each learning example~\cite{ 2010_nips_y.noh_generative}. 
The whole parameters of these kinds of the algorithm are prohibitively huge when the number of examples becomes large.

In our approach, we follow~\cite{2012_nips_j.wang_parametric} to learn a local metric for a part of the feature space of documents, in which case the number of learned metrics $m$ can be considerably smaller than $n$, the size of the examples collection.

Suppose we have learned $m$ local metrics $\left\{ \mathbf{M}_{1}, \ldots, \mathbf{M}_{m}\right\}$ and the associated anchor points $\left\{ p_{1}, \ldots p_{r}, \ldots p_{m}\right\}$. 
The choice of anchor points and the computation of local metrics are described in Subsection~\ref{subsec:Computation-of-Basis}. 
Then the similarity model $f(q,p)$ between two documents $p_{i}$ and $p_{j}$ is extended from Eq.~(\ref{eq:global_m_distance}) as follows:

\begin{eqnarray} 
f(p_{i},p_{j}) & = & d_{\mathbf{M}(p_{i})}(p_{i}, p_{j}) \\
\mathbf{M}(p_{i}) & = & \sum_{r=1}^{m}w_{r}(p_{i}) \mathbf{M}_{r},
\label{eq:local_metrics}
\end{eqnarray}

where $w_{r}(p_{i})\ge0$ is the weight of document $p_{i}$ for local metric $\mathbf{M}_{r}$. 
The PSD constraints of $\mathbf{M}(p)$ is automatically satisfied if all local metric $\mathbf{M}_{r}$ are PSD matrices. 
These formulation includes $m$ anchor documents and $p_{i}$ should be close to these anchor documents~\cite{2011_pami_d.ramanan_local}. 
It is clear that the \textit{ideal candidate document} should be close to several high relevant documents. 
Therefore, we can employ these high relevant documents as anchor documents to construct the local metric space around the \textit{ideal candidate document}.

With the above assumption and observation, the task of information retrieval precedes in the following steps: 

\begin{enumerate} 
	\item Given a candidate collection $\mathcal{D}_{q}$ for query $q$ , we employ high/low relevant documents to compute a $\mathbf{M}$ and find a anchor point $p_{r}$ to maximize the ranking scorer under the metric $\mathbf{M}$ by computing $\left(p_{i}-p_{r}\right) \mathbf{M} \left(p_{i}-p_{r} \right)^{\top}$. 
	\item After sampling $m$ candidate collection to find $m$ anchor documents and $m$ associated metrics, we can construct the \textit{ideal candidate document} based on a combination of $m$ anchor documents. 
	\item We can build an evaluation function to measure the similarity between candidate document and \textit{ideal candidate document}, then, sort these documents via the similarity to \textit{ideal candidate document}. 
\end{enumerate}

\subsection{Computation of Basis Metrics\label{subsec:Computation-of-Basis}}

Before constructing the local metrics in Eq.~(\ref{eq:local_metrics}), we need to learn $m$ local metrics. 
With the assumption that each local metric $\mathbf{M}_{r}$ represents a part of feature space, we can employ the classical single metric learning algorithm associated with a subset of the triplets from a part of examples space.

In this paper, we extend the state-of-the-art global metric learning algorithm GMML~\cite{ 2016_icml_p.zadeh_geometric} into local metric forms. 
The extension consists of two parts: 

\begin{enumerate} 
	\item The local basis metric associated with the triplets set $\mathcal{D}_{r}$ is computed by the original GMML. 
	\item The smooth weighting function $w_{r}(p)$ is computed from 
	Eq.~(\ref{eq:smooth_weight_function}). 
\end{enumerate} 

Given a subset of the triplets $\mathcal{T}_{r} = \left(p_{i}, p_{j}, p_{k}\right)$ such that $p_{i}$ is more similar to $p_{j}$ than to $p_{k}$, we can extract the similarity set $S_{r}$ and the dissimilarity set $D_{r}$ by following the instruction in Section~\ref{subsec:Sampling-Strategy}. Then we construct two corresponding matrices:

\begin{eqnarray} 
\mathbf{S}_{r} & = & \sum_{(p_{i},p_{j})\in S_{r}}\left(p_{i}-p_{j}\right)\left(p_{i}-p_{j}\right)^{\top}\\
\mathbf{D}_{r} & = & \sum_{(p_{i},p_{k})\in D_{r}}\left(p_{i}-p_{k}\right)\left(p_{i}-p_{k}\right)^{\top} 
\end{eqnarray}

Then, the basic optimization formulation of local metric $\mathbf{M}_{r}$ is defined as follows:

\begin{equation} 
\underset{\mathbf{M}_{r}\succ0}{\min}\quad 
h(\mathbf{M}_{r}):=\text{tr}\left(\mathbf{M}_{r}\mathbf{S}_{r}\right) + \text{tr}\left(\mathbf{M}_{r}^{-1}\mathbf{D}_{r}\right)
\label{eq:objective_gmml} 
\end{equation}

Equation (\ref{eq:objective_gmml}) implies that GMML will return a single local metric $\mathbf{M}_{r}$ that minimize the sum of distances over all the similar pairs $S_{r}$ and maximize the distance over all the dissimilar pairs $D_{r}$.

The closed-form solution of Eq.~(\ref{eq:objective_gmml}) is obtained by

\begin{equation} 
\nabla h(\mathbf{M}_{r}) = \mathbf{S}_{r} - \mathbf{M}_{r}^{-1} \mathbf{D}_{r} \mathbf{M_{r}^{-1}} 
\end{equation}

Taking $\nabla h(\mathbf{M}_{r})=0$, we obtain:

\begin{equation}
\mathbf{M}_{r}\mathbf{S}_{r}\mathbf{M}_{r}=\mathbf{D}_{r}
\label{eq:riccati_equation} 
\end{equation}

Equation (\ref{eq:riccati_equation}) is a Riccati equation whose unique
solution is~\cite{2016_icml_p.zadeh_geometric}

\begin{equation}
\mathbf{M}_{r}=\mathbf{S}_{r}^{-1/2}\left(\mathbf{S}_{r}^{1/2}\mathbf{D}_{r}\mathbf{S}_{r}^{1/2}\right)^{1/2}\mathbf{S}_{r}^{-1/2}
\label{eq:solution_gmml}
\end{equation}

In experiments, $\mathbf{M}_{r}$ is efficiently computed from Cholesky-Schur method~\cite{ 2016_nlaa_b.iannazzo_geometric}.

\subsection{Smoothing Weight Functions}

Lots of researchers try to provide the insights of their local metric learning approaches~\cite{ 2012_nips_j.wang_parametric, 2012_nips_s.hauberg_geometric} by modeling their methods from the perspective of Riemannian metric. 
An important property about the Riemannian metric is that a Riemannian metric $M(p)$ on a manifold $\mathcal{M}$ is a smoothly varying inner product $\left\langle x_{i},x_{j} \right\rangle_{p} = x_{i}^{T} \mathbf{M}(p)x_{j}$ in the tangent space $\mathcal{T}_{p}\mathcal{M}$ of each point $p\in\mathcal{M}$. 
From Lemma 1 in~\cite{2012_nips_s.hauberg_geometric}, when the weight function $w_{r}(p)$ is smooth with $p$, Eq.~(\ref{eq:local_metrics}) will be a well-studied Riemannian metric. 
Therefore, any well-designed local metric methods should provide a smooth weight function.

Another important issue is that the weight function $w_{r}(p)$ should reflect the fitness of the local metric $\mathbf{M}_{r}$. 
Suppose $(p,p_{r})\in S_{r}$, it indicates that $\mathbf{M}_{r}$ is the best local metric to measure the similarity between $p_{r}$ and other examples, which means that Eq.~(\ref{eq:riccati_equation}) should be robust against the additive similar pair $(p,p_{r})$. 
Therefore, the weight function $w_{r}(p)$ should be in the opposite to $\mathbf{M}_{r}(p-p_{r})(p-p_{r})^{T}\mathbf{M}_{r}$.

Take the limit as an example, if $\mathbf{M}_{r}(p-p_{r})(p-p_{r})^{T} \mathbf{M}_{r} = \mathbf{0}$, then,

\begin{equation}
\mathbf{M}_{r}\left(\mathbf{S}_{r}+(p-p_{r})\right)\left(\mathbf{S}_{r}+(p-p_{r})\right)^{T}\mathbf{M}_{r}=\mathbf{D}_{r}
\label{eq:new_input_riccati_equation} 
\end{equation}

The solution of Eq.~(\ref{eq:new_input_riccati_equation}) is the same with Eq.~(\ref{eq:solution_gmml}), which indicates that $\mathbf{M}_{r} (p-p_{r}) (p-p_{r})^{T} \mathbf{M}_{r}$ is a proper measurement whether the $\mathbf{M}_{r}$ is the optimal local metric for the document $p$.

By taking consideration about the above observation, we propose the smoothing weight functions~\cite{1988_cs_d.broomhead_multivariable} as:

\begin{equation} 
w_{r}(p)=\exp\left(-\frac{\rho}{2}\left\Vert
p-p_{r}\right\Vert _{\mathbf{M}_{r}}\right), 
\label{eq:smooth_weight_function}
\end{equation}

where, $\left\Vert \cdot\right\Vert _{\mathbf{M}_{r}}^{2}$ is the L2 norm with the metric $\mathbf{M}_{r}$.

\begin{equation} 
\left\Vert p-p_{r}\right\Vert_{\mathbf{M}_{r}}^{2} = \text{tr}\left(\mathbf{M}_{r}(p-p_{r})(p-p_{r})^{T}\mathbf{M}_{r}\right)
\label{eq:trace_norm} 
\end{equation}

From Eq.~(\ref{eq:trace_norm}) and Eq.~(\ref{eq:new_input_riccati_equation}), we can easily know $\left\Vert p-p_{r}\right\Vert _{\mathbf{M}_{r}}$ is a proper measurement about the similarity between query $p$ and the anchor point $p_{r}$ associated with the local metric $\mathbf{M}_{r}$.

Therefore, our evaluation function is formulated as:

\begin{equation}
f_{q}(p,\Phi_{q})=-\sum_{r=1}^{m}\Phi_{q}^{(r)}\cdot\exp\left(-\left\Vert
p-p_{r}\right\Vert _{M_{r}}\right)\cdot\left\Vert p-p_{r}\right\Vert
_{M_{r}},
\label{eq:evaluation_function} 
\end{equation}

where, $\Phi_{q}\in\mathbb{R}^{m}$, $\Phi_{q}^{(r)} = \exp\left( \rho_{q}^{(r)}/2 \right)$ is the key parameter we need to learn in order to find a better manifold structure. 
Higher $f_{q}(p,\Phi_{q})$ means that $p$ is closer to the \textit{ideal candidate document}. 
In the next subsection, we will introduce our exploration to optimize $\Phi$ for a specific ranking problem.

\subsection{Update of $\Phi$\label{subsec:Update-of-Phi}}

In the above subsection, we formulate a general local metric framework in Eq.~(\ref{eq:evaluation_function}) to represent the manifold structure. 
From the theoretical analysis in~\cite{2011_pami_d.ramanan_local}, the whole space of $\Phi$ keeps the learned manifold smooth. 
Therefore, we define our loss function under the popular Weighted Approximate Rank Pairwise (WARP) framework~\cite{2010_ml_j.weston_large} and optimize the associated objective function to obtain an optimal solution for ranking task.

The WARP loss for a given set of candidate document $\mathcal{D}_{q}$ with
query ID $q$ is defined as:

\begin{equation}
\mathcal{L}\left(q\right)=\frac{1}{\left|\mathcal{D}_{q}^{+}\right|}\sum_{p\in\mathcal{D}_{q}^{+}}L\left(v_{q}\left(p^{+}\right)\right),
\label{eq:warp_loss} 
\end{equation}

where $v_{q}(p^{+})$ is the number of violators in $\mathcal{D}_{q}$ for positive $p^{+}$, defined as:

\begin{equation}
v_{q}(p^{+})=\sum_{p^{-}\in\mathcal{D}_{q}^{-}}\mathbf{I}\left[f_{q}\left(p^{-},\Phi_{q}\right)-f_{q}\left(p^{+},\Phi_{q}\right)\right] 
\end{equation}

To obtain better NDCG score, $L\left(\cdot\right)$ is defined as:

\begin{equation} 
L(k)=\sum_{i=1}^{k}\frac{1}{\log_{2}\left(i+1\right)}
\end{equation}

In order to optimize $\Phi_{q}$, we follows the methods in~\cite{ 2010_ml_j.weston_large, 2014_icml_d.lim_efficient} to approximate $L\left(v_{q}(p^{+})\right)$ by a continuous formulation with hinge loss:

\begin{equation}
\sum_{p^{-}\in\mathcal{V}_{q,p^{+}}}L\left(\left|\mathcal{V}_{q,p^{+}}\right|\right)\frac{\left[\zeta-f_{q}\left(p^{+},\Phi_{q}\right)+f_{q}\left(p^{-},\Phi_{q}\right)\right]_{+}}{\left|\mathcal{V}_{q,p^{+}}\right|},
\label{eq:formulation_hinge_loss} 
\end{equation}

where for a given $q,p^{+}$, $\zeta$ is the hinge loss margin.
$\mathcal{V}_{q,p^{+}}$ is the set of violators with hinge loss:

\begin{equation} 
\mathcal{V}_{q,p^{+}}=\left\{ p^{-}\in\mathcal{X}_{q}^{-}\ |\
f_{q}(p^{+},\Phi_{q})\right\} 
\end{equation}

In order to obtain an unbiased estimation of the loss function in Eq.~(\ref{eq:formulation_hinge_loss}), we can randomly sample $q$, $p^{+} \in \mathcal{D}_{q}$ and find an violator $p^{-}$ such that $\zeta + f_{q} \left(p^{-}, \Phi_{q}\right) > f_{q} \left(p^{+},\Phi_{q}\right)$. 
In this situation, the tuple of $\left(q,p^{+},p^{-}\right)$ has the following contribution to Eq.~(\ref{eq:formulation_hinge_loss}):

\begin{equation}
l\left(q,p^{+},p^{-}\right)=L\left(\left|\mathcal{V}_{q,p^{+}}\right|\right)\left(\zeta-f_{q}\left(p^{+},\Phi_{q}\right)+f_{q}\left(p^{-},\Phi_{q}\right)\right)
\label{eq:loss_for_triple} 
\end{equation}

From the WARP framework, $\left|\mathcal{V}_{q,p^{+}}\right|$ can be approximated by $\left\lfloor \left|\mathcal{D}_{q}^{-}\right| / N_{q}\right\rfloor $, where $N_{q}$ is the number of less relevant documents $p^{-}$ drawn with replacement from $\mathcal{D}_{q}^{-}$ until a violator is found.

Finally, the stochastic gradient descent for the parameter $\Phi$ can be easily conducted at iteration $t$ as:

\begin{align} 
& \Phi_{q}(t+1) \nonumber\\ 
= & \Phi_{q}(t)-\mu\frac{\partial
	l\left(q,p^{+},p^{-}\right)}{\partial\Phi_{q}(t)},\\ 
= & \Phi_{q}(t)-\mu
L\left(\left\lfloor \frac{\left|\mathcal{D}_{q}^{-}\right|}{N_{q}}\right\rfloor
\right)\cdot\left[\frac{\partial
	f_{q}(p^{-},\Phi_{q}(t))}{\partial\Phi_{q}(t)}-\frac{\partial	
	f_{q}\left(p^{+},\Phi_{q}(t)\right)}{\partial\Phi_{q}(t)}\right],\label{eq:update_phi} 
\end{align}

where $\frac{\partial f_{q}(p,\Phi_{q})}{\partial\Phi_{q}} = \left[\frac{\partial f_{q} \left(p, \Phi_{q}^{(r)}\right)} {\partial\Phi_{q}^{(r)}} \right]_{r=1 \ldots m}$. 
To avoid over-fitting, we project $\Phi_{q}^{(r)}$ to zero when Eq.~(\ref{eq:update_phi}) leads to negative value. 
We take derivation from Eq.~(\ref{eq:evaluation_function}) to obtain:

\begin{equation}
\frac{\partial f_{q}\left(p,\Phi_{q}^{(r)}\right)}{\partial\Phi_{q}^{(r)}}=\exp\left(-\left\Vert p-p_{r}\right\Vert _{M_{r}}\right)\cdot\left\Vert p-p_{r}\right\Vert _{M_{r}} 
\end{equation}

Overall, our proposed algorithm is illustrated in Figure \ref{fig:framework} and summarized in Algorithm \ref{alg:lgmml}.

\subsection{Sampling Strategy\label{subsec:Sampling-Strategy}}

Our approach will not iterate all triplets for $\mathcal{D}$ introduced in Section \ref{subsec:Computation-of-Basis}, because learning the global ranking model from all triplets is an NP-hard problem~\cite{2012_thesis_b.mcfee_more}.
Hence, we choose to stochastically sample the triplets $\left(p_{i}, p_{j}, p_{k} \right)$ from candidate documents. 
$p_{i}$ and $p_{j}$ representing similar documents are sampled from high relevant documents, then $p_{k}$ is sampled from the less relevant documents. 
In our implementation, we only sample $p_{k}$ from the documents with zero relevant label.

For sampling procedure in Section~\ref{subsec:Update-of-Phi}, $\Phi_{i}$ and $\Phi_{j}$ are independent for two queries $i$ and $j$. 
Therefore, the update can be computed in a highly parallel way.

\begin{algorithm} 
	\SetAlgoNoLine
	\KwIn{ $\mathcal{D}^{+}$ : positive set of documents, $\mathcal{D}^{-}$ :
		negative set of documents, $\lambda$ : regularization parameter
	}
	\KwOut{$M\in\mathbb{S}_{+}^{d}$ : Mahalanobis metric;}
	$\mathbf{S}=\lambda\mathbf{I}+\sum_{p_{i}\neq p_{j},p_{i}\in\mathcal{D}^{+},p_{j}\in\mathcal{D}^{+}}\left(p_{i}-p_{j}\right)\left(p_{i}-p_{j}\right)^{\top}$;

	$\mathbf{D}=\lambda\mathbf{I}+\sum_{p_{i}\in\mathcal{D}^{+},p_{j}\in\mathcal{D}^{-}}\left(p_{i}-p_{j}\right)\left(p_{i}-p_{j}\right)^{\top}$;
	
	$\mathbf{M}=\mathbf{S}^{-1/2}\left(\mathbf{S}^{1/2}\mathbf{D}\mathbf{S}^{1/2}\right)^{1/2}\mathbf{S}^{-1/2}$
	
	\caption{Geometric Mean Metric Learning (GMML)~\cite{ 2016_icml_p.zadeh_geometric}}
	
	\label{alg:gmml} 
\end{algorithm}

\begin{algorithm} 
	\SetAlgoNoLine
	
	\KwIn{Candidate set for $c$ queries $\left\{
		\mathcal{D}_{1},\mathcal{D}_{2},\ldots,\mathcal{D}_{q},\ldots,\mathcal{D}_{c}\right\} $, $m$ : number of local metrics, $T$ : number of iteration, $\mu$ : step size, $\zeta$: hinge loss margin	
	}
	
	\KwOut{$\left\{
		\left(p_{1},M_{1}\right),\left(p_{2},M_{2}\right)\ldots,\left(p_{m},M_{m}\right)\right\} $ : set of local metrics and associated anchor points, $p\in\mathbb{R}^{d}$, $M\in\mathbb{S}_{+}^{d}$, $\Phi\in\mathbb{R}^{c\times m}$ : weights for local metrics to model the \textit{ideal candidate documents} for each queries	
	}
	
	\For{$q\in\left[1,c\right]$}{
		Extract $\mathcal{D}_{q}^{+}$ and $\mathcal{D}_{q}^{-}$ from
		$\mathcal{D}_{q}$;	
	}

	\For{$i\in\left[1,m\right]$}{	
		Sample $\mathcal{D}_{i}^{+}$ and $\mathcal{D}_{i}^{-}$ from $\left\{
		\mathcal{D}_{q}\right\} _{q\in\left[1,c\right]}$;
		
		$M_{i}$ = $\text{GMML}\left(\mathcal{D}_{i}^{+},\mathcal{D}_{i}^{-}\right)$;
		
		\For{$p\in\mathcal{D}_{i}^{+}$}{	
			$\Gamma_{p}^{(i)}\leftarrow$Sort $\mathcal{D}_{i}$ in ascending order by
			computing $\left\Vert p-d\right\Vert _{M_{i}}^{2}\ \forall
			d\in\mathcal{D}_{q}$;	
		}
		
		Find the anchor point $p_{r}$ with maximum NDCG score of
		$\Gamma_{p_{r}}^{(i)}$;
		
	}
	
	
	\For{$t=1$ $\text{ to }$ $T$}{
		
		Sample a tuple $\left(q,p^{+},p^{-}\right)$ from $\left\{
		\mathcal{D}_{q}\right\} _{q\in\left[1,c\right]}$ such that

		$\zeta+f_{q}\left(p^{+},\Phi_{q}\left(t\right)\right)>f_{q}\left(p^{-},\Phi_{q}\left(t\right)\right)$;
		
		$N_{q}\leftarrow$ the number of less relevant documents drawn with
		replacement from $\mathcal{D}_{q}^{-}$ until $p^{-}$ is found;
		
		$\Phi_{q}\left(t+1\right)=\left[\Phi_{q}\left(t\right)-\mu
		L\left(\left\lfloor
		\frac{\left|\mathcal{D}_{q}^{-}\right|}{N_{q}}\right\rfloor
		\right)\cdot\left[\frac{\partial
			f_{q}(p^{-},\Phi_{q}(t))}{\partial\Phi_{q}(t)}-\frac{\partial
			f_{q}\left(p^{+},\Phi_{q}(t)\right)}{\partial\Phi_{q}(t)}\right]\right]_{+}$;
	}		
	\caption{L-GMML to Rank}
	\label{alg:lgmml} 
\end{algorithm}

\section{Evaluation}

In this section, we discuss the implementation of our approach for the LtR problem and display extensive experiments evaluating our methodology in comparison to the state-of-the-art (R-MLR, GBRT, and $\lambda$-MART). 
Our design on experiments tackle the following questions:

\begin{itemize}
	\item Do we develop a correct localized extension to the global GMML? 
	To answer this question, we generate varied scale synthetic datasets to evaluate the performance gain against global metric learning algorithm when a different number of local metrics invoke in our L-GMML approach to prove the correctness. 
	\item Is our assumption on the existence of local structures reasonable? 
	If reasonable, does our solution enjoy high computational efficiency and the good scalability for scaled datasets? 
	We make comparisons with the state-of-the-art metric learning algorithms for ranking in the query-dependent model on scaled datasets. 
	We attempt to demonstrate the improvements of our approach over other metric learning algorithms. 
	\item Does our LtR algorithm have any amazing properties? 
	We conduct experiments on real-world large-scale datasets to illustrate the enormous improvement of our approach on accuracy compared with the dominant ranking methods like GBRT and $\lambda$-MART in the query-independent framework.
\end{itemize}

\subsection{Experiments Setting}

In our experiments, we have implemented our local GMML (L-GMML) algorithm in Julia\footnote{http://julialang.org/}, the source code is released at Github\footnote{https://github.com/yxsu/LtR.jl}. 
To make a fair comparison against the state-of-the-art ranking methods, we also implement R-MLR, GBRT and $\lambda$-MART in Julia. 
We take RankLib\footnote{http://sourceforge.net/p/lemur/wiki/RankLib/}, an open-source implementation of the GBRT and $\lambda$-MART algorithms and
MLR\footnote{https://github.com/bmcfee/mlr} as references to implement these
algorithms in Julia.

Our program is executed on an Ubuntu 14.04 LTS server with 12 Intel Xeon E5-2620 cores and 128GB main memory. 
All baseline methods and our method are performed in a parallel way to fully utilize the computational resources. 
Our R-MLR implementation is based on the parallel MLR-ADMM~\cite{ 2013_icml_d.lim_robust}. 
GBRT and $\lambda$-MART come from RankLib.

The statistical tests in the following experiments are computed over the values for Mean Average Precision (MAP) and Normalized Discounted Cumulative Gain
(NDCG)~\cite{2002_tois_k.jarvelin_cumulated} at the top $k$ retrieved documents denoted as NDCG@$k$. 
These two metrics are the most important and frequently used in information retrieval community to evaluate a given permutation of a ranked list using binary and multi-relevance order.

\subsection{Datasets}

For all real-world datasets, we split each of them into two components: 1), the training set is used to learn ranking models; 
2), the test set is purely used to evaluate the performance of the learned ranking models.

All the datasets we use are freely available online for scientific purpose.
Such datasets can be divided into two groups:

\subsubsection{Query-dependent Dataset}

We employ CAL10K~\cite{ 2010_mir_d.tingle_exploring} to make fair comparisons between our L-GMML and R-MLR. 
Because, in the original paper, R-MLR performs well on the CAL10K dataset. 
Following the experiments in~\cite{2013_icml_d.lim_robust}, we use a subset of the CAL10K dataset, which is provided as ten 40/30/30 splits of a collection of 5419 songs.

\begin{table} 
	\caption{Different kinds of song representation}
	\centering
	
	\begin{tabular}{ccc} 
		\toprule & \# of features & \# of songs \\ 
		\midrule Audio & 1,024 & 5,419\\ 
		Lyrics-128 & 128 & 2000\\ 
		Lyrics-256 & 256 & 2000\\
		\bottomrule 
	\end{tabular}	
	\label{table:info_songs} 
\end{table}

\subsubsection{Query-Independent Datasets}

In this subsection, we employ two popular real-world large-scale datasets: Yahoo! and MSN to evaluate the competitive performance of proposed L-GMML against the state-of-the-art query-independent LtR methods.

Yahoo! datasets come from Yahoo! Learning to Rank Challenge~\cite{ 2011_jmlr_o.chapelle_yahoo}. 
The datasets consist of feature vectors extracted from query-url pairs along with relevance judgment labels.

In our experiments, we also employ the two set of MSN learning to rank\footnote{https://www.microsoft.com/en-us/research/project/mslr/} datasets: MSLR-10K and MSLS-30K, both of which consists of 136 features extracted from query-url pairs. 
The MSN datasets provide relevance judgment labels ranging from 0 (irrelevant) to 4 (perfect match). 
In experiments, each MSN dataset is partitioned into five subsets for five-fold cross validation.

The complete statistical information about these datasets are listed at Table \ref{table:info_datasets}.

\begin{table*}[t] 
	\caption{Characteristics of publicly available large-scale datasets for learning to rank}
	
	\centering
	
	\begin{tabular}{cccccccccccl} 
		\toprule Name & \multicolumn{3}{c}{\# of Queries} & \multicolumn{3}{c}{\# of Doc.} & Rel. Levels & \# of Features & Year \\
		\cmidrule(r){2-4}  \cmidrule(r){5-7} & Train & Vali. & Test & Train &
		Vali. & Test \\
		\midrule 
		Yahoo! Set I & 19,944 & 2,994 & 6,983 & 473,134 &
		71,083 & 165,660 & 5 & 519 & 2010 \\
		Yahoo! Set II & 1,266 & 1,266 & 3,798 &
		34,815 & 34,881 & 103,174 & 5 & 596 & 2010 \\
		MSLR-WEB10K & 6,000 & 2,000 & 2,000 & 723,412 & 235,259 & 235,259 & 5 & 136 & 2010 \\
		MSLR-WEB30K & 31,531 & 6,306 & 6,306 & 3,771k & 6,306 & 753k & 5 & 136 & 2010 \\
		\bottomrule
	\end{tabular}
	
	\label{table:info_datasets} 
\end{table*}

\subsection{Evaluation of the Proposed Approach}

In our L-GMML model, the most important hyper-parameter is the number of local metrics, which has significant influence on the overall model performance. 
We will evaluate the correction of our localized extension method from synthetic datasets, and reveal the impact of the metric numbers.

\subsubsection{Global GMML vs Local GMML}

In this subsection, we attempt to employ multi-class classification problem to verify the correction of the local metric learning algorithm. 
Because multi-class synthetic datasets certainly contain local structures around the center of each class.
If the accuracy gain is observed, we can also address the objective that local metric learning approach is designed to extend the global metric learning method's ability of modeling complex data manifold.

We employ the normal distribution to generate synthetic datasets with multiple centers and 95\% confidence interval. 
The datasets with \{10,50,100\} classes are denoted as Synthetic-10, Synthetic-50, Synthetic-100 respectively. 
In these synthetic datasets, we assign the index of class to the relevant label of the corresponding data point.

\begin{figure*}	
\centering
\scalebox{0.8}{
	\begin{tikzpicture}
	\begin{groupplot}[
	group style={horizontal sep=35pt, group size=4 by 1}, 
	height=5cm, 
	xlabel={\# of metrics}]
	\nextgroupplot[
	title=Synthetic-2, 
	ylabel={MAP}] 
	\addplot[mark=x] table[
	x=num_local_metrics, 
	y=local_auc]
	{global_vs_local_synthetic_2_class.dat}; 
	\addplot table[
	x=num_local_metrics, 
	y=global_auc]
	{global_vs_local_synthetic_2_class.dat}; 
	\addlegendentry{L-GMML}
	\addlegendentry{GMML} 
	\nextgroupplot[
	title=Synthetic-10, 
	ylabel={MAP}, 
	legend style={at={(1,0.6)}}] 
	\addplot[mark=x] table[
	x=num_local_metrics, 
	y=local_auc]
	{global_vs_local_synthetic_10_class.dat}; 
	\addplot table[
	x=num_local_metrics, 
	y=global_auc]
	{global_vs_local_synthetic_10_class.dat}; 
	\addlegendentry{L-GMML}
	\addlegendentry{GMML} 
	\nextgroupplot[
	title=Synthetic-50, 
	ylabel={MAP}, 
	legend style={at={(1,0.5)}}] 
	\addplot[mark=x] table[
	x=num_local_metrics, 
	y=local_auc]
	{global_vs_local_synthetic_50_class.dat}; 
	\addplot table[
	x=num_local_metrics, 
	y=global_auc]
	{global_vs_local_synthetic_50_class.dat}; 
	\addlegendentry{L-GMML}
	\addlegendentry{GMML} 
	\nextgroupplot[
	title=Synthetic-100, 
	ylabel={MAP}, 
	legend style={at={(1,0.4)}}] 
	\addplot[mark=x] table[
	x=num_local_metrics, 
	y=local_auc]
	{global_vs_local_synthetic_100_class.dat}; 
	\addplot table[
	x=num_local_metrics, 
	y=global_auc]
	{global_vs_local_synthetic_100_class.dat}; 
	\addlegendentry{L-GMML}
	\addlegendentry{GMML} 
	\end{groupplot} 
	\end{tikzpicture}
}
\caption{Comparisons between global GMML and local GMML on synthetic datasets. The performance is measured by MAP}

\label{fig:global-vs-local-synthetic}
\end{figure*}
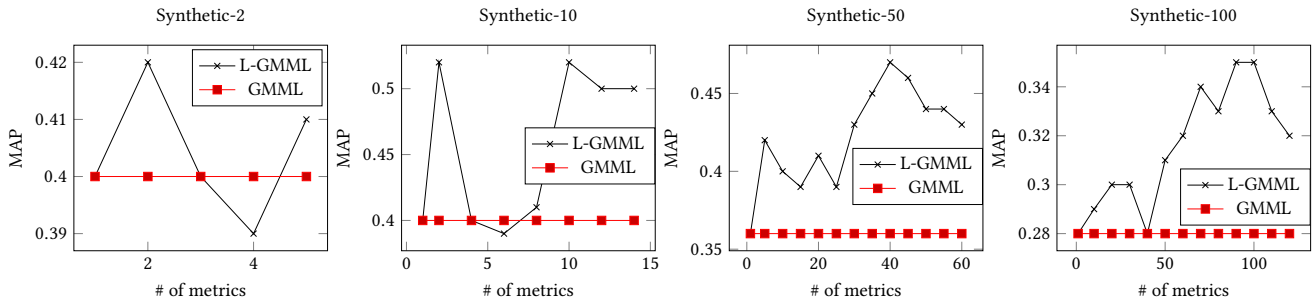

We report the performance gain of the proposed local GMML against the global GMML in Figure \ref{fig:global-vs-local-synthetic}. 
We can easily find the fact that when the number of local metrics is approximate to the number of the real centers in Gaussian synthetic data distribution, the relative accuracy gain of local metrics is maximized. 
This observation meets the objective of local metric learning approach.

\subsubsection{The Number of Local Metrics}

In this subsection, we will evaluate the significance of the number of local metrics, which is typically the most important parameter in the field of local metric learning. 
A large number of local metrics will enhance the algorithm's ability to model the complex manifold structure. 
However the computational complexity increases linearly with the number of local metrics. 
In experiments, we need to carefully tune the number of local metrics to make the balance between model's ability and computational complexity.

\begin{figure*} 
\centering 
\begin{tikzpicture} \begin{groupplot}[group
style={horizontal sep=35pt, vertical sep=40pt, group size=4 by 2}, height=4cm,
xlabel={\# of metrics}] \nextgroupplot[title=Audio, ylabel={MAP}] \addplot
table[x=num_local_metrics, y=auc] {num_local_metrics_audio.dat};
\nextgroupplot[title=Audio with PCA, ylabel={MAP}] \addplot
table[x=num_local_metrics, y=auc] {num_local_metrics_audio_pca.dat};
\nextgroupplot[title=Lyrics128, ylabel={MAP}] \addplot
table[x=num_local_metrics, y=auc] {num_local_metrics_lyrics_128.dat};
\nextgroupplot[title=Lyrics256, ylabel={MAP}] \addplot
table[x=num_local_metrics, y=auc] {num_local_metrics_lyrics_256.dat};
\nextgroupplot[title=Yahoo Set I, ylabel={NDCD@10}] \addplot
table[x=num_local_metrics, y=ndcg] {num_local_metrics_yahoo_1.dat};
\nextgroupplot[title=Yahoo Set II, ylabel={NDCD@10}] \addplot
table[x=num_local_metrics, y=ndcg] {num_local_metrics_yahoo_2.dat};
\nextgroupplot[title=MSLR-10K, ylabel={NDCD@10}] \addplot
table[x=num_local_metrics, y=ndcg] {num_local_metrics_mslr_10.dat};
\nextgroupplot[title=MSLR-30K, ylabel={NDCD@10}] \addplot
table[x=num_local_metrics, y=ndcg] {num_local_metrics_mslr_30.dat};
\end{groupplot} \end{tikzpicture}
\caption{The variation of performance caused by different number of local
	metrics}
\label{fig:different_num_local_metrics} 
\end{figure*}
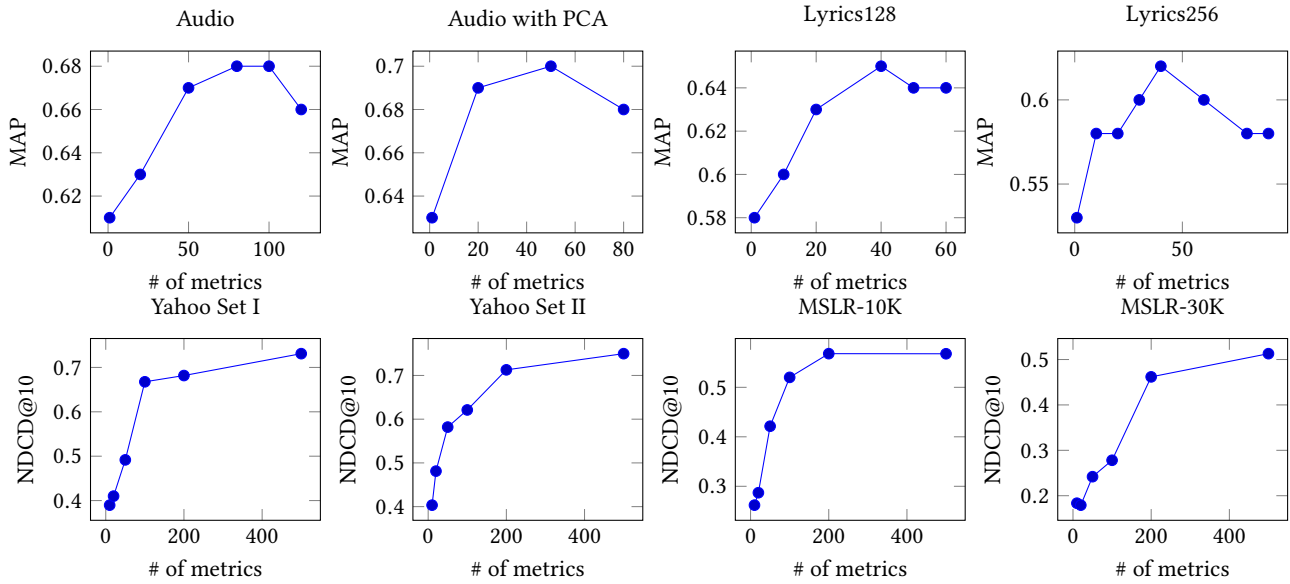

Figure \ref{fig:different_num_local_metrics} displays the impact of the number of local metrics on all datasets used in our paper. 
For all datasets, localized method can compete with the corresponding global method with a single metric.
This fact proves that our localized extension is reasonable. 
Another obvious observation is that the optimal number of local metrics varies dramatically among different datasets, since it is decided by the complexity of the manifold structure sealed in the data space.

\subsubsection{Scalability}

In our experiments, the synthetic datasets is primarily invoked to evaluate the scalability of our approach.

\begin{figure} 
\centering
\scalebox{0.9}{
\begin{tikzpicture}
\begin{semilogyaxis}[xlabel={\# of dimension}, ylabel={Training time (s)},
legend style={at={(1, 0.1)}, anchor=east}] \addplot table[x=dimension, y=rmlr]
{training_time_different_scaled.dat}; \addplot table[x=dimension, y=lgmml]
{training_time_different_scaled.dat}; \legend{R-MLR, L-GMML}
\end{semilogyaxis} \end{tikzpicture}
}
\caption{Training time of L-GMML on different scaled synthetic datasets}
\label{fig:training_time_different_scaled}
\end{figure}
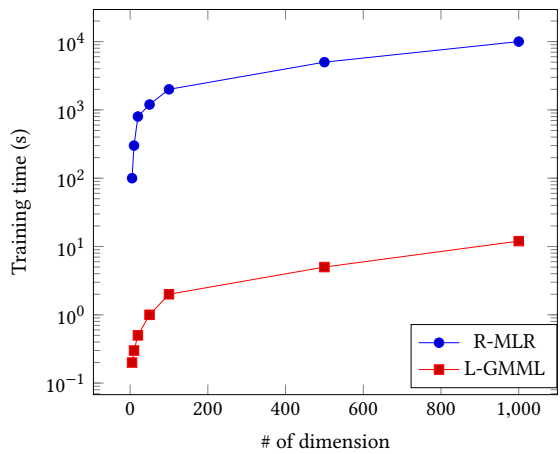

Due to the limited scalability of real-world datasets, we synthesize datasets with the feature dimensionality scaled from 10 to 1000. 
In this experiment, we fix the number of local metrics as 10 since we only concern about the computational complexity on different scaled dimensions instead of the optimal number of local metrics. 
Figure \ref{fig:training_time_different_scaled} illustrates the training time of our L-GMML on these datasets.

Compared with other local metric learning methods, the less training time come from two-fold issues: (1) the GMML in Algorithm \ref{alg:gmml} is very fast.
(2) The update of weighting function in our approach is relatively simple and straightforward.
It does not involve the huge computational resources to find the optimal form.

\subsection{Comparison with R-MLR}

The Robust Metric-Learning-to-Rank (R-MLR)~\cite{2013_icml_d.lim_robust} is the most competitive metric learning method for ranking. 
It retrieves relevant examples in response to a query instance. 
To make direct comparisons, we need to modify our approach by assigning all anchor points to the query instance. 
Because our approach is originally designed for the query-independent framework.

In this set of experiments, we evaluate our approach on the music similarity task, because the R-MLR method is verified to be successful in music similarity task compared with other metric-learning-to-rank methods such as MLR~\cite{ 2010_icml_b.mcfee_metric}, $L1$-MLR~\cite{2006_kdd_r.rosales_learning}. 
For each song $p_{i}$, a relevant set $\mathcal{D}_{i}^{+} \subset \mathcal{D}_{\text{train}}$ is defined as the subset of songs in the training set performed by the top 10 most similar artists to the performer of $p_{i}$, where the similarity between artists is measured by the number of shared users in a sample of collaborative filter data~\cite{ 2012_itaslp_b.mcfee_learning}. 
This top-10 thresholding results in the relevant sets in this data being asymmetric and non-transitive.
Therefore, the traditional pairwise metric learning methods do not work in this situation. 
However, our approach based on the sampling on the relevant set is not necessary to obey the symmetric and transitive properties.

The experiments are conducted on two different kinds of song representation: audio and lyrics, whose details are listed in Table \ref{table:info_songs}. 
We use recommended candidate hyper-parameters in the original paper to tune R-MLR on validation set and select the best parameter to evaluate the performance of the model.

Since the scalability of the original R-MLR is limited, the experiments of R-MLR employ the latent features compressed by PCA. 
Our approach has no such problem and is suitable to conduct the training process on the original 1,024 features.

\begin{table} 
	\caption{Comparison on the training time of R-MLR and L-GMML. The
		number of local metrics in L-GMML is fixed as 50}
	
	\centering
	
	\begin{tabular}{ccc} \toprule Time (s) & R-MLR & L-GMML\\ \midrule Audio & N/A
		& 38 \\ Audio with PCA & 607 & 4.7 \\ Lyrics-128 & 302 & 2.6 \\ Lyrics-256 &
		1241 & 7.8 \\ \bottomrule 
	\end{tabular} 
\end{table}

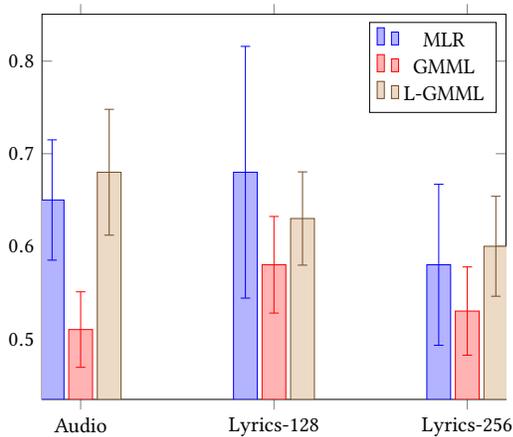
\begin{figure}
	
\centering 
\scalebox{0.9}{
\begin{tikzpicture}[baseline] \begin{axis}[ybar,
xtick=data, symbolic x coords={Audio, Lyrics-128, Lyrics-256},]
\addplot+[error bars/.cd, y dir=both, y explicit relative,] table[x=dataset,
y=auc_RMLR, y error=err_RMLR] {music_similarity_performance_auc.dat};
\addplot+[error bars/.cd, y dir=both, y explicit relative,] table[x=dataset,
y=auc_GMML, y error=err_GMML] {music_similarity_performance_auc.dat};
\addplot+[error bars/.cd, y dir=both, y explicit relative,] table[x=dataset,
y=auc_LGMML, y error=err_LGMML] {music_similarity_performance_auc.dat};
\addlegendentry{MLR} \addlegendentry{GMML} \addlegendentry{L-GMML} \end{axis}
\end{tikzpicture}
}

\caption{Music similarity performance of each algorithm on the three feature representation Audio, Lyrics-128 and Lyrics-256. 
Performance was measured by MAP and averaged across 10 folds.}

\label{fig:auc_music_similar_three_methods} 
\end{figure}

Figure \ref{fig:auc_music_similar_three_methods} illustrates the performance of three metric learning algorithms. 
We fix the number of local metrics in our L-GMML as 1 to obtain the global GMML algorithm. 
The motivation of making such comparison is that we attempt to demonstrate the different influence of the new GMML algorithm and our proposed L-GMML algorithm on the performance improvements.

Therefore, we can draw the conclusion from the experiments in this subsection that the proposed approach outperforms other metric learning algorithms for the ranking problem regarding accuracy and computational efficiency.

\subsection{Comparisons on Large-scale Real-world Datasets}

We attempt to find amazing features of our method in the comparisons with two state-of-the-art ranking methods, Gradient-Boosted Regression Trees (GBRT) and $\lambda$-MART on Yahoo! Set I\&II, MSLR-10K, and MSLR-30K. 
Because they have been proved to be the most effective in the Yahoo! learning to rank challenge and become the dominant methods in the LtR field.

\begin{table*} 
\caption{Performance of GBRT, $\lambda$-MART and proposed L-GMML on large-scale real-world datasets. 
Results of MSLR-WEB10K and MSLR-WEB30K are averaged from the 5 folds in the datasets.}

\label{table:performance_gbrt_mart_lgmml}

\centering

\begin{tabular}{cccccccccc} 
	\toprule & & \multicolumn{2}{c}{GBRT} &
	\multicolumn{2}{c}{$\lambda$-MART} & \multicolumn{2}{c}{L-GMML} \\
	\cmidrule(r){3-4} \cmidrule(r){5-6} \cmidrule(r){7-8}
	\multicolumn{2}{c}{Dataset} & Test Set & Time (min.) & Test Set & Time (min.)
	& Test Set & Time (min.) \\ \midrule
	
	& NDCG@5 & 0.6529 & 41.2 & 0.6567 & 46.5 & \textbf{0.6698} & 28.1 \\ Yahoo!
	Set I & NDCG@10 & 0.6824 & 43.3 & \textbf{0.7060} & 48.0 & 0.6715 & 28.9 \\ &
	NDCG@20 & 0.6912 & 41.5 & \textbf{0.7091} & 46.9 & 0.6934 & 28.8 \\
	
	\cmidrule(r){1-8} & NDCG@5 & 0.6731 & 37.6 & 0.6791 & 43.1 & \textbf{0.7096}
	& 26.5 \\ Yahoo! Set II & NDCG@10 & 0.6817 & 36.8 & 0.7062 & 43.3 &
	\textbf{0.7264} & 26.6 \\ & NDCG@20 & 0.6954 & 37.4 & 0.7087 & 43.8 &
	\textbf{0.7219} & 26.4 \\
	
	\cmidrule(r){1-8} & NDCG@5 & 0.4019 $\pm$ \textbf{0.0083} & 49.4 $\pm$ 5.2 &
	0.4417 $\pm$ 0.0131 & 58.3 $\pm$ 2.8 & \textbf{0.4771} $\pm$ 0.0951 & 19.7
	$\pm$ 2.1 \\ MSLR-WEB10K & NDCG@10 & 0.4342 $\pm$ 0.0219 & 48.3 $\pm$ 2.1 &
	0.4513 $\pm$ \textbf{0.0196} & 57.6 $\pm$ 3.8 & \textbf{0.5390} $\pm$ 0.0812
	& 19 $\pm$ 3.1 \\ & NDCG@20 & 0.4512 $\pm$ 0.0279 & 48.8 $\pm$ 3.8 & 0.4634
	$\pm$ \textbf{0.0257} & 57.1 $\pm$ 5.2 & \textbf{0.551} $\pm$ 0.0728 & 19
	$\pm$ 2.8 \\
	
	\cmidrule(r){1-8} & NDCG@5 & 0.409 $\pm$ 0.0312 & 167 $\pm$ 28.6 & 0.3812
	$\pm$ \textbf{0.0297} & 182 $\pm$ 19.8 & \textbf{0.4837} $\pm$ 0.0715 & 71.7
	$\pm$ 2.7 \\ MSLR-WEB30K & NDCG@10 & 0.4146 $\pm$ 0.0327 & 177 $\pm$ 30.1 &
	0.409 $\pm$ \textbf{0.0232} & 183 $\pm$ 17.9 & \textbf{0.4976} $\pm$ 0.0619 &
	71.9 $\pm$ 3.9 \\ & NDCG@20 & 0.421 $\pm$ 0.361 & 167 $\pm$ 27.3 & 0.4112
	$\pm$ \textbf{0.0240} & 181 $\pm$ 10.7 & \textbf{0.5038} $\pm$ 0.0718 & 72.5
	$\pm$ 5.3 \\ \bottomrule 
\end{tabular} 
\end{table*}

For these four datasets, the feature domain varies dramatically. 
To avoid for challenging the floating point precision in complex mathematical computation, we preprocess these four datasets by normalizing each feature dimension with 2-norm. 
For the stochastic sampling procedure in Algorithm \ref{alg:lgmml}, to find the optimal model, we try different initial weight values $\Phi(1)$ ranging from $0.1$ to $10$, the hinge loss margin $\zeta$ ranging from $0.01$ to $1$.

\begin{table} 
\caption{Per-document scoring time of GBRT, $\lambda$-MART and L-GMML on Yahoo! and MSLR datasets. The scoring time is united in $ms$}
\label{table:score_time_per_document}
\centering

\begin{tabular}{cccc} \toprule & GBRT & $\lambda$-MART & L-GMML\\ \midrule
	Yahoo! Set I & 276 & 302 & 421\\ Yahoo! Set II & 218 & 286 & 421 \\
	MSLR-WEB10K & 73 & 92 & 158 \\ \bottomrule 
\end{tabular} 
\end{table}

The training time of GBRT and $\lambda$-MART is sensitive to the number of trees in both of the models. 
The number of local metrics also determines the training time of L-GMML. 
When we plan to make comparisons on the accuracy and training time of three methods, we fix the number of trees of GBRT and $\lambda$-MART as 5000 and the number of local metrics as 500. 
The motivation of these choices is that the performance of these two methods become stable on the four datasets. 
The comprehensive comparisons on a different measurement of the above three methods are illustrated in Table \ref{table:performance_gbrt_mart_lgmml}. 
From the table, we can draw a conclusion that our approach enjoys a huge advantage in accuracy compared with the state-of-the-art ranking methods.

Currently, the only disadvantage of our approach lies in scoring time. 
Table \ref{table:score_time_per_document} displays the comparisons about the time of scoring documents. 
Our algorithm heavily relies on the scoring for each document in different stages, which is less efficient than GBRT and $\lambda$-MART. 
On the other hand, our approach is simple in structure, and GMML in the first stage is also efficient.
Therefore, our method still has an advantage in computationally efficiency. The time-consuming comparison in Table \ref{table:performance_gbrt_mart_lgmml} can prove this statement.

\section{Conclusion}

In this paper, we focus on improving the accuracy of LtR methods by utilizing the local structure of documents and degrading irrelevant features. 
We firstly developed a localized GMML algorithm for the query-independent ranking framework. 
Specifically, we proposed a concept called \textit{ideal candidate document} to adopt metric learning for ranking algorithm from a query-dependent model to widely used query-independent model.
In our approach, a well defined smooth weighting function is optimized by reducing the popular WARP loss, which is defined for the candidate document set of a given query. 
Then we can efficiently score document by calculating the distance between candidate documents and a nonexistent \textit{ideal candidate document} from an optimized metric space. 
The experiments prove that our approach outperforms both of the state-of-the-art query-dependent algorithms and query-independent algorithms.

\section{Acknowledgments}

We would like to thank the anonymous reviewers for their valuable comments and suggestions to improve the quality of the paper. 
The work described in this paper was partially supported by the Research Grants Council of the Hong Kong Special Administrative Region, China (No. CUHK 14208815 and No. CUHK 14234416 of the General Research Fund), and 2015 Microsoft Research Asia Collaborative Research Program (Project No. FY16- RES-THEME-005).

\bibliographystyle{ACM-Reference-Format} 
\bibliography{library}

\end{document}